\documentclass[a5paper, 11pt]{article}
\usepackage{graphicx}
\usepackage{fullpage}
\usepackage{fancyhdr}
\usepackage{enumerate}
\usepackage{mathtools}
\usepackage{amsmath}
\usepackage{txfonts}

\usepackage[utf8]{inputenc}
\usepackage[english]{babel}
\usepackage{amssymb}
\usepackage[bottom=1.5cm,top=2.5cm,left=2.1cm,right=2.1cm]{geometry}
\usepackage{sectsty}
\usepackage[symbol]{footmisc}
\sectionfont{\large}
\addto\captionsswedish{}

\begin{document}
\pagestyle{fancy}
\renewcommand{\headrulewidth}{0pt}
\fancyhf{}
\cfoot{\thepage}
\thispagestyle{empty}
\clearpage
\pagenumbering{arabic} 

\Large
\noindent
\begin{center}
\textbf {Chemonuclear Transmutation \\and Noble Metal Synthesis}\footnote[2]{This article is an excerpt from Hidetsugu Ikegami’s book manuscript.} 
\end{center}
\vspace{1.5em}
\small
\noindent
\begin{center}
\text{Hidetsugu Ikegami}\\
\vspace{0.25em}
\footnotesize
\text{1930--2025}
\end{center}
\footnotesize
\noindent
\begin{center}
\text{Professor Emeritus, Osaka University}
\text{Formerly Director, Research Center for Nuclear Physics, Osaka University}
\text{Honorary Doctorate, Uppsala University}
\end{center}

\noindent
\begin{center}
\textit{Edited by}\\
\small
\text{Masako Ikegami}
\end{center}
\footnotesize
\noindent
\begin{center}
\text{Professor Emeritus, Institute of Science Tokyo/Tokyo Institute of Technology}
\text{Cooperative Researcher, Research Center for Nuclear Physics, Osaka University}
\text{E-mail: mikegami@rcnp.osaka-u.ac.jp}
\end{center}

\newpage
\setcounter{page}{31}
\large
\noindent
\begin{center}
\text{\textbf{2.} Chemonuclear Transmutation and Noble Metal Synthesis}
\end{center}

\small
\noindent
\begin{center}
\text{Hidetsugu Ikegami and Masako Ikegami}
\end{center}\Large

\Large
\normalsize
\hyphenchar\font=-1
\section*{}
\begin{abstract}
\noindent 
In the system of metallike hydride--electron donor mixtures, adsorbed hydrogen atoms are transformed into metallic states and alloyed. Dense itinerary s-electrons supplied by the electron donor cause a drop in the melting point of metallic hydrogen. As a result, this system reveals thermodynamical liquid activity. The coherent $\mathrm{D_2}$-$\mathrm{D_2}$ fusion and the $\mathrm{D_3}$-$\mathrm{D_3}$ chemonuclear fusion are enhanced with factors of $10^{20}\sim10^{30}$ and $10^{30}\sim10^{46}$ respectively at $T$=460 K, hence the production of intense He ions of 23.8 MeV kinetic energy. The ions or $\alpha$-particles induce enhanced cascade chemonuclear reactions towards diverse and useful element synthesis. This phenomenon is in the nature of the Big Bang nucleosynthesis that takes place in an inhomogeneous universe. An application of the $\alpha$-induced chemonuclear reaction -- the transmutation of $^{90}$Sr and $^{137}$Cs -- is prescribed in this chapter. This system of metallike hydride--electron donor mixtures raises the possibility of radioactive waste vanishment. Stimulated by $\alpha$-particle irradiation, trans-gold nuclei undergo exothermic $\alpha$-cluster emissions, producing coherently line-up $\alpha$-clusters. In the chemonuclear D-D fusion, the 4$\alpha$-cluster reveals the thermodynamical activity of oxygen atoms. In the hydrogen-Ni nanopowder-Li mixtures, dispersively charged trans-gold atoms undergo enhanced chemonuclear $\alpha$-cluster emissions, hence the mass synthesis of noble metals. The case of chemonuclear pion production is presented at the end of this chapter.
\vspace{.5em}

\noindent
\it{Keywords:} cascade chemonuclear reaction; chemonuclear fission; element synthesis; pion production; waste disintegration
\end{abstract}

\noindent
\normalsize
\hyphenchar\font=-1
\section*{2.1 Enormous Fusion Enhancement}

In the system of metallike hydride--electron donor mixtures, the H-H chemonuclear fusion produces He ions of 23.8 MeV energy. This fusion reaction is enhanced with factors over $10^{20}$. As such, the H-H chemonuclear fusion is in the nature of enhanced pycnonuclear fusion in condensed plasma in supernova progenitors [1-3]. Moreover, this fusion reaction may induce enhanced cascade chemonuclear reactions that are fairly similar to the Big Bang nucleosynthesis [4-6].

For the nuclear fusion between colliding reactants $\mathrm{X_1\,and\,X_2}$ that form a compound nucleus M
\begin{equation}
\mathrm{X_1 + X_2 \longrightarrow M \, } \label{eq1}
\end{equation}
the reaction cross section [4-6] is
$$ \sigma = \frac{S(E)}{E^{1/2} (E + E_{\mathrm{S}})^{1/2}} \mathrm{exp} \left[ -\pi \sqrt{\frac{E_{\mathrm{G}}}{E + E_{\mathrm{S}}}} \, \right] $$
\begin{equation}
\sim \frac{S(E)}{E} \mathrm{exp} \left[ -\pi \sqrt{\frac{E_{\mathrm{G}}}{E}} \, \right]  \label{eq2}
\end{equation}
assuming that 
\begin{equation}
\mathrm{Relative\,kinetic\, energy}\: E\: \gg\: \mathrm{Screening\, energy\,}\: {E_\mathrm{s}\, \mathrm{between\, the\, reactants}} \label{eq3}
\end{equation}

In Eq. (2), the exponential decay factor is the Coulomb penetration factor, and the cross-section factor $S(E)$ is a quantity intrinsic to the nuclear fusion. The Gamov energy $E_{\mathrm{G}}$ is given as
\begin{equation}
E_{\mathrm{G}} = \frac{Z_1 Z_2 e^2}{4 \pi \epsilon_0 r^*_{12}} = \frac{0.0992 (Z_1 Z_2)^2 A_1 A_2 }{A_1 + A_2} \, \mathrm{MeV}  \label{eq4}
\end{equation}
Here $r^*_{12}$ denotes the nuclear Bohr radius of the colliding reactants.

The reaction cross section of Eq. (\ref{eq2}) is effectively magnified by the enhancement factor [7-15], as detailed in Chapter 1.
\begin{equation}
K = \mathrm{exp} \left[ -\frac{\Delta \textit{G}_r}{k_{\mathrm{B}} T} \, \right] \sim \mathrm{exp} \left[ \frac{\Delta \mathrm{\phi^{*}_r}}{k_{\mathrm{B}} T} \, \right]  \label{eq5}
\end{equation}
where $K$ corresponds to the equilibrium constant of the atomic fusion reaction united with the nuclear fusion; $k_{\mathrm{B}}$ and $T$ denote the Boltzmann constant and the temperature of the liquid, respectively. 

The Gibbs free energy change $\Delta G_{\mathrm{r}}$ is given by the formation energy $\Delta G_{\mathrm{f}}$, and the chemical potential change $\mathrm{\Delta \phi^*_r}$ in the atomic fusion is given by the chemical potential $\mathrm{\phi^*}$ of the reactant- and product atoms, hence
\begin{equation}
\Delta G_{\mathrm{r}} = \Delta G_{\mathrm{f}} (\mathrm{M}) - \Delta G_{\mathrm{f}} (\mathrm{X}_1) - \Delta G_{\mathrm{f}} (\mathrm{X_2}) \,  \label{eq6}
\end{equation}
or
\begin{equation}
\Delta \phi^*_\mathrm{r} = \Delta \phi^* (\mathrm{M}) - \Delta \phi^* (\mathrm{X}_1) - \phi^* (\mathrm{X_2}) \,  \label{eq7}
\end{equation}
Here the formation energy $\Delta G_{\mathrm{f}}$ is empirically related to the Pauling electronegativity $\chi$ [16, 17, 18] as follows:
\begin{equation}
-\Delta G_{\mathrm{f}}=2.5\chi \label{eq8}
\end{equation}
Table 2.1 shows the formation Gibbs energy deduced from Eq. (8) and the chemical potential $\phi^*$ observed for atoms implanted in Be metal [16, 19, 20].

Although this enhancement is infeasible in gas plasma like the solar interior, it is common among spontaneous reactions in liquids, regardless of the kind of liquid. Similar enormous enhancement was observed in the slow ion-induced chemonuclear fusion in metallic Li liquids [8-15]. The rate enhancement is similar to that of the enhanced pycnonuclear reaction in metallic-hydrogen liquid plasma. Widom's “thermodynamical activity of liquids” may explain this fact [21] as the bulk/collective features of spontaneous reactions caused by thermodynamic force in liquids consisting of reactant particles. This thermodynamic force is subject to the chemical potential change in the reactions. This thermodynamic relationship is independent of the kind of reactant particle and the nature of the microscopic interparticle interactions [22].

\newpage
\hyphenchar\font=-1
\small
\noindent Table 2.1 \, Formation Gibbs energy $-\Delta G_{\mathrm{f}}$ and chemical potential $\phi^*$.

\small
\begin{center}

\begin{tabular}{| c | c | c | c | c | c | c | c | c |} \hline
Z & 3 & 4 & 5 & 6 & 7 & 8 & 9 & 11 \\ \hline
Element & Li & Be & B & C & N & O & F & Na \\ \hline
$-\Delta G_{\mathrm{f}}$ &2.45& 3.93 & 5.10 & 6.38 & 7.60 & 8.60 & 9.95 & 2.33 \\ \hline
$\phi^*$&2.85&4.20&4.80$$&6.23$$&7.00$$& / & / & 2.70 \\ \hline
\end{tabular}

\begin{tabular}{| c | c | c | c | c | c | c | c | c |c |} \hline
12 & 13 & 14 & 15 & 16 & 17 & 19 & 20 & 21 & 22 \\ \hline
Mg & Al & Si & P & S & Cl & K & Ca & Sc & Ti \\ \hline
3.28 & 4.03 & 4.75 & 5.48 & 6.45 & 7.90 & 2.05 & 2.50 & 3.40 & 3.85 \\ \hline
3.45 & 4.20 & 4.70 & / & / & / & 2.25 & 2.55 & 3.25 & 3.65\\ \hline
\end{tabular}

\begin{tabular}{| c | c | c | c | c | c | c | c | c |c |} \hline
23 & 24 & 25 & 26 & 27 & 28 & 29 & 30 & 31 & 32 \\ \hline
V & Cr & Mn & Fe & Co & Ni & Cu & Zn & Ga & Ge \\ \hline
4.08 & 4.15 & 3.88 & 4.58 & 4.70 & 4.78 & 4.75 & 4.13 & 4.53 & 5.03 \\ \hline
4.25 & 4.65 & 4.45 & 4.93 & 5.10 & 5.20 & 4.55 & 4.10 & 4.10 & 4.55 \\ \hline
\end{tabular}

\begin{tabular}{| c | c | c | c | c | c | c | c | c | c |} \hline
33 & 34 & 35 & 36 & 37 & 38 & 39 & 40 & 41 & 42 \\ \hline
As & Sc & Br & Kr & Rb & Sr & Y & Zr & Nb & Mo \\ \hline
5.45 & 6.38 & 7.40 & 7.25 & 2.05 & 2.38 & 3.05 & 3.33 & 4.00 & 5.40 \\ \hline
4.80 & / & / & / & 2.10 & 2.40 & 3.20 & 3.40 & 4.00 & 4.65 \\ \hline
\end{tabular}

\begin{tabular}{| c | c | c | c | c | c | c | c | c | c |} \hline
43 & 44 & 45 & 46 & 47 & 48 & 49 & 50 & 51 & 52 \\ \hline
Tc & Ru & Rh & Pd & Ag & Cd & In & Sn & Sb & Te \\ \hline
4.75 & 5.50 & 5.70 & 5.50 & 4.83 & 4.23 & 4.45 & 4.90 & 5.13 & 5.25 \\ \hline
5.30 & 5.40 & 5.40 & 5.45 & 4.45 & 4.05 & 3.90 & 4.15 & 4.40 & / \\ \hline
\end{tabular}

\begin{tabular}{| c | c | c | c | c | c | c | c | c | c |} \hline
53 & 54 & 55 & 56 & 57 & 72 & 73 & 74 & 75 & 76 \\ \hline
I & Xe & Cs & Ba & La & Hf & Ta & W & Re & Os \\ \hline
6.65 & 6.50 & 1.98 & 2.23 & 2.75 & 3.25 & 3.75 & 5.90 & 4.75 & 5.50  \\ \hline
/ & 4.18 & 1.95 & 2.32 & 3.05 & 3.55 & 4.05 & 4.80 & 5.40 & 5.40 \\ \hline
\end{tabular}
\begin{tabular}{| c | c | c | c | c | c | c | c | c | c | c | c |} \hline
77 & 78 & 79 & 80 & 81 & 82 \\ \hline
Ir & Pt & Au & Hg & Tl & Pb \\ \hline
5.50 & 5.70 & 6.35 & 5.00 & 5.10 & 5.83 \\ \hline
5.55 & 5.65 & 5.15 & 4.20 & 3.90 & 4.10 \\ \hline
\end{tabular}
\begin{tabular}{| c | c | c | c | c | c | c | c | c | c | c | c |} \hline
83 & 90 & 91 & 92 & 93 & 94 \\ \hline
Bi & Th & Pa & U & Np & Pu\\ \hline
5.05 & 3.25 & 3.75 & 3.45 & 3.40 & 3.20\\ \hline
4.15 & 3.30 & / & 4.05 & / & 3.80\\ \hline
\end{tabular}
\end{center}

\newpage
\small
\begin{center}

\begin{tabular}{| c | c | c | c | c | c | c |} \hline
Z & 58 & 59 & 60 & 61 & 62 & 63 \\ \hline
Element & Ce & Pr & Nd & Pm & Sm & Eu \\ \hline
$-\Delta G_{\mathrm{f}}$ & 2.80 & 2.83 & 2.85 & 3.00 & 2.93 & 3.00 \\ \hline
$\phi^*$ & / & / & / & / & / & / \\ \hline
\end{tabular}

\begin{tabular}{| c | c | c | c | c | c | c | c |} \hline
64 & 65 & 66 & 67 & 68 & 69 & 70 & 71\\ \hline
Gd & Tb & Dy & Ho & Er & Tm & Yb & Lu\\ \hline
3.00 & 3.00 & 3.05 & 3.08 & 3.10 & 3.13 & 2.75 & 3.18\\ \hline
/ & / & / & / & / & / & / & /\\ \hline
\end{tabular}
\end{center}

\begin{figure}[h]
\centering
\includegraphics[scale=0.45, angle=90]{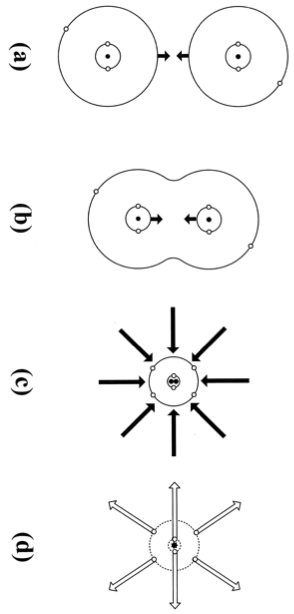}
\end{figure}
\begin{center}
\small
Fig. 2.1 \, Evolution of the $^7$Li-$^7$Li chemonuclear fusion.
\end{center}

\noindent 
\normalsize
Figure 2.1 shows the evolution of the $^7$Li-$^7$Li chemonuclear fusion through the formation of an intermediate united atom (i.e., a quasi-C atom) induced by a $^7$Li atom with an acceleration energy $\sim$ 1 MeV in the Li liquid. We highlight the following experimental issues:

\begin{enumerate}[1.]
\item \textit{Twin colliding $^7$Li atoms with a relative kinetic energy of $E$=0.5 MeV in the metallic Li liquid} 

At this energy level, a nuclear collision takes place but electrons (open circles) in the K- and L-shell orbitals adjust their electronic states continuously and smoothly to the nuclear collision process, because the electrons gyrate considerably faster than the colliding atoms' speed. The de Broglie wave length of itinerary s-electrons in the metallic Li liquid covers the space of tens Li atoms, hence the collective activity of the liquid. This liquid, consisting of Li ions and itinerary s-electrons, is the macroscopically correlated solvent that reacts with those traversing solute atoms/ions in the united atomic and nuclear fusion.

\item \textit{Intermediate diatomic molecule Li$_2$ formation} 

In the first phase of the collision process, the colliding atoms tend towards diatomic molecule Li$_2$ formation at the minimal point of the chemical potential. This chemical potential, however, is not the minimum point for those atoms colliding with the energy $E$=0.5 MeV. They are directed to form a new dense intermediate complex -- i.e., a quasi-C atom -- at their colliding or turning point.

\item \textit{Quasi-C atom formation alongside the formation of an ultradense intermediate $^7$Li-$^7$Li nuclear complex (twin closed circles)} 

The colliding Li atoms collapse alongside quasi-C atom formation, resulting in a sharp reduction in their atomic volume by a factor of 0.065 [12]. The thermodynamic force specified by the chemical potential -- i.e., $\Delta G_\mathrm{f}$(Li) or  $\phi^* (\mathrm{Li})$ and $\Delta G_\mathrm{f}$(C) or $\phi^* (\mathrm{C})$ from Table 2.1 -- dominates the united atomic and nuclear fusion, as seen in Eqs. (\ref{eq6})-(\ref{eq8}). The collision time characterized by the period of the zero point oscillation of colliding nuclei at their classical turning point -- i.e., $\tau_0 \sim 1\times 10^{-21}\, \mathrm{s}$ (see Eq. (3) in [12]) -- is very short. In the presence of thermodynamic force (shown by octaval arrows), the collision time is prolonged at the melting point of Li metal (460 K) by a factor of $\mathrm{exp}(-\Delta G_\mathrm{r}/ k_{\mathrm{B}} T) = 10^{16}\,\mathrm{or}\,\, \mathrm{exp}\ (\Delta \phi^*/k_{\mathrm{B}} T) = 10^6$, hence the formation of a metastable united atom or a quasi-C atom. Inside the quasi-C atom, a pair of colliding nuclei are confined in the sphere of zero-point oscillation $7.4\times 10^{-5} \, (\mathrm{pm})^3$ (see Eq. (2) in [12]), forming themselves into an ultradense intermediate nuclear complex. The density of this complex is $3\times 10^{14}\, \mathrm{kg \, m^{-3}}$, i.e., a billion times the solar interior density (around $10^5 \, \mathrm{kg \, m^{-3}}$). The density of the white dwarf progenitors of supernovae is around $10^{13} \, \mathrm{kg\,m^{-3}}$. As such, a quasi-C atom formed in the metallic Li liquid is a packing unit of ultradense nuclear complex immersed in the macroscopically coherent sea of itinerary s-electrons. 

\item \textit{Nuclear fusion releasing spectator neutrons via a highly excited intermediate nucleus $^{12}\mathrm{C}^*$} 

The rate of the nuclear fusion is enhanced by a factor of $10^6\sim 10^{16}$ and most of the colliding Li ions undergo nuclear fusion.
\end{enumerate}

Radiative nuclear capture reactions are rare due to the momentum mismatch between some MeV reactant particles and outgoing photons. As a result, the cross sections are some $\mu$b at most [23]. Small electronegativity elements, i.e., alkaline and alkaline earth metals, can mostly take part in the enhanced $\alpha$-capturing chemonuclear reaction. In these elements, the power of an atom to attract electrons to itself (i.e., the chemical potential $\phi ^*$) is small (see Table 2.1). For these elements, the $\alpha$-capturing chemonuclear reaction is specified by the chemical potential change

\begin{equation}
\Delta \phi^{*}_r = \phi ^*(Z+2)-\phi^*(Z) > 0 \label{eq9}\,
\end{equation}

\noindent in the reaction and will be enhanced by the factor $K$ in Eq. (5).

\hyphenchar\font=-1
\section*{2.2 Cascade Chemonuclear Reactions}

The isomeric transitions following the nuclear capture reactions is connected with internal or external electron conversion. It should be noted that the conversion electron coefficient is some hundredths at most depending on the transition energy, the multi-pole order of transitions and the atomic number of electron conversion atoms. In the following examples, we assume that the reactants are metals or hydrides/deuterides reduced to metals on the surface of such hydrogen adsorbent as Ni and Pd nanocrystal clusters.

In the system of metallike hydride--dense electron donor mixtures, adsorbed hydrogen atoms are transformed into metallic hydrogen [20] around the octahedral sites of Ni or Pd lattices [24, 25]. Dense itinerant s-electrons supplied by the donor cause a drop in the melting point of the metallic hydrogen and induce their liquefaction. In the presence of thermodynamical liquid activity, hydrogen atoms become strongly correlated with the surrounding bulk of atoms, hence macroscopic scale collectivity. The coherent $\mathrm{D_2}$-$\mathrm{D_2}$ and $\mathrm{D_3}$-$\mathrm{D_3}$ fusion reactions at $T=460$ K -- enhanced with factors over $10^{20}$$\sim10^{30}$ and $10^{30}$$\sim10^{46}$ respectively -- take place, producing $\alpha$-particles of 23.8 MeV kinetic energy. The $\alpha$-particles reduce their speed when traversing the medium. Note that the speed of $\alpha$-particles is below 0.1 $c$ (where $c$ is the speed of light). The value of 0.1 $c$ is close to the 2s-orbital electron gyration speed of the Ca atoms, but this value is below the 1s orbital electron speed. As the Ca atoms interact with $\alpha$-particles, the electrons adjust their configuration smoothly and continuously to match the Ca+$\alpha$ nuclear reaction, hence the chemonuclear He-induced reaction enhanced with a factor of $\mathrm{exp[\Delta \phi^*/k_B}T]$ at temperature $T$. In the presence of dense s-electrons, the chemical potential change may result in enormous enhancement, hence the cascade chemonuclear reaction [26]. The important role of high-density itinerant s-electrons was shown by experiment [26, 27].

The example of $\gamma$-ray missing radiative $\alpha$-capture reactions of the Ca isotopes is given below:
\begin{equation}
\mathrm{Ca} + \alpha \rightarrow \mathrm{Ti^{2+}} + \mathrm{e}^-_\mathrm{ext} \, , \mathrm{Q} > 0\,
\end{equation}
\begin{equation}
\mathrm{Ca} + \alpha \rightarrow \mathrm{Ti^{3+}} + \mathrm{e}^-_\mathrm{int}\,
\end{equation}
\begin{equation}
\mathrm{Ca} + \alpha \rightarrow \mathrm{Ti^{4+}} + \mathrm{e_{}}^-_\mathrm{int} + \mathrm{e_A^-}\,
\end{equation}
Here $\mathrm{e_{ext}^-}$, $\mathrm{e_{int}^-}$, and $\mathrm{e_{A}^-}$ denote external conversion, internal conversion, and nuclear charge shake-off Auger electron, respectively.

If the Ca atoms are surrounded by such heavier atoms as the Pd and Ni atoms, Eq. (10) will be the most likely case. This reaction is specified by the chemical potential change
\begin{align*}
\Delta \phi^*_r = \phi^*(\mathrm{Ti})-\phi^*(\mathrm{Ca})-\Delta \mathrm{G_f(Ti^{2+}_{liq})}
\end{align*}
\begin{equation}
= 3.65 - 2.55 + 2.75 = 3.85\:\mathrm{eV}
\end{equation}
Here the chemical potential $\phi^*$ is obtained from the alloy data (Table 2.1). As for the formation Gibbs energy of $\mathrm{Ti^{2+}_{liq}}$ ion, we use $\mathrm{\Delta G_f(Ti^{2+}_{aq})}$ in the chemical data [18]. In a system revealing thermodynamical liquid activity, the capture reaction of Eq. (10) is enhanced with the factor $K$ in Eq. (14) as
\begin{equation}
\mathrm{log} \, K=42 \,\,\,\, \mathrm{at} \,T=460\:\mathrm{K}\,
\end{equation}
This means there is no space for the $\gamma$-ray emitting $\alpha$-capture reaction. On the other hand, the chemonuclear $\alpha$-capture reactions are often associated with such large-scale secondary soft photons as soft x-rays and phonons generated by the conversion electrons. This kind of soft x-rays was observed in the D-D chemonuclear fusion experiments [26]. 

As shown by Eqs. (10)-(12), the Ti ions that have been produced are neutralized quickly in the presence of electron donors in the system and will undergo a successive reaction
\begin{equation}
\mathrm{Ti+\alpha} \rightarrow \mathrm{Cr^{2+}+e^-_{ext}} \, , \, \mathrm{Q} > 0\,
\end{equation}
specified by the chemical potential change
\begin{align*}
\Delta \phi^*_r = \phi^*(\mathrm{Cr}) - \phi^*\mathrm{(Ti)} - \Delta \mathrm{G_f(Cr^{2+}_{liq})}\,
\end{align*}
\begin{equation}
= 4.65 - 3.65 + 1.49 = 2.49\:\mathrm{eV}\,
\end{equation}
This reaction is enhanced with the factor
\begin{equation}
\mathrm{log \, }K = 27 \, \mathrm{at} \, T=\mathrm{460\:K}\,
\end{equation}

In the case of reactant $^{40}$Ca (abundance 97\%), the cascade $\alpha$-capture reaction and $\beta$-disintegration are as follows:
\begin{align*}
^{40}\mathrm{Ca} + \alpha \rightarrow \, ^{44}\mathrm{Ti} \, (48\mathrm{y}) \, , \, \mathrm{Q} = \mathrm{5.13\,MeV}
\end{align*}
\begin{equation}
^{44}\mathrm{Ti} + \alpha \rightarrow \, ^{48}\mathrm{Cr} \, (23\mathrm{h}) \, , \, \mathrm{Q} = \mathrm{7.69\,MeV}
\end{equation}
\begin{align*}
^{48}\mathrm{Cr(Q=1.4\,MeV)}\,  \xrightarrow[]{\mathrm{EC}}  \, ^{48}\mathrm{V(16d, \, Q = 4.01\,MeV)} \xrightarrow[]{\beta^+} \mathrm{^{48}Ti}
\end{align*}
We find the accumulation of $^{48}$Ti that has been shown by experiment [26]. The $^{48}$Ti found by experiment is mostly due to the cascade transmutation of Eq. (18) rather than the cascade $^{44}$Ca + d $\rightarrow ^{46}$Sc, $^{46}$Sc + d $\rightarrow ^{48}$Ti reactions, because the abundance of $^{44}$Ca is only 2\%. In the reaction $^{46}$Sc + d, the enhancement is log $K$ = 4 at $T$=460 K, which is far below that of Eq. (17).

In the Ca + $\alpha$ reaction, Sr hydrides/deuterides are reduced to metallic Sr on the surface of such adsorbent clusters as Ni or Pd nanocrystal powders, hence the chemonuclear reaction
\begin{equation}
\begin{split}
^{88}\mathrm{Sr} + \alpha \rightarrow \, ^{92}\mathrm{Zr}^{3+} + \mathrm{e_{int}^-} \, , \, \mathrm{Q = 2.97\,MeV} \\
^{90}\mathrm{Sr} + \alpha \rightarrow \, ^{94}\mathrm{Zr}^{3+} + \mathrm{e_{int}^-} \, , \, \mathrm{Q = 3.76\,MeV}\,
\end{split}
\end{equation}
specified by the chemical potential change
\begin{align*}
\Delta \phi_r^* = \phi^*(\mathrm{Zr})-\Delta \mathrm{G}_f(\mathrm{Zr^{3+}_{liq}})-\phi^*\mathrm{(Sr)}
\end{align*}
\begin{equation}
= 3.40 + 1.76 - 2.40 = 2.76\:\mathrm{eV}
\end{equation}
and enhanced with the factor
\begin{equation}
\mathrm{log\,}K=30 \, \, \, \mathrm{at} \, T=\mathrm{460\:K}
\end{equation}
The $^{92,94}$Zr$^{3+}$ ions that have been produced are quickly neutralized and will undergo a successive reaction as follows:
\begin{equation}
\begin{split}
^{92}\mathrm{Zr} + \alpha \rightarrow \, ^{96}\mathrm{Mo}^{3+} + \mathrm{e_{int}^-} \, , \, \mathrm{Q = 2.76\,MeV}
\\
^{94}\mathrm{Zr} + \alpha \rightarrow \, ^{98}\mathrm{Mo}^{3+} + \mathrm{e_{int}^-} \, , \, \mathrm{Q = 3.27\,MeV}
\end{split}
\end{equation}
\begin{align*}
\Delta \phi_r^* = \phi^*(\mathrm{Mo})-\Delta \mathrm{G}_f(\mathrm{Mo^{3+}_{liq}})-\phi^*\mathrm{(Zr)}
\end{align*}
\begin{equation}
=4.65 + 1.00 - 3.40 = 2.25\:\mathrm{eV}
\end{equation}
with the enhancement factor
\begin{equation}
\mathrm{log\,}K\sim24 \, \, \, \mathrm{at} \, T=\mathrm{460\:K}
\end{equation}
The transmutation of Sr towards the production of Mo was observed [27]. A part of the Mo atoms will undergo a successive reaction that produces Ru; however, its enhancement is much smaller than that of Eq. (24).

The $^{133}$Cs atoms in the metallic form on the surface of adsorbent nanocrystal clusters will undergo the chemonuclear reaction
\begin{equation}
^{133}\mathrm{Cs} + \alpha \rightarrow \, ^{137}\mathrm{La}^{3+} + \mathrm{e_{int}^-} \, , \, \mathrm{Q = 1.58\,MeV}\,
\end{equation}
specified by the chemical potential change
\begin{align*}
\Delta \phi_r^*(\mathrm{La^{3+}}) = \phi^*(\mathrm{La})- \phi^*(\mathrm{Cs}) - \Delta \mathrm{G}_f(\mathrm{La^{3+}_{liq}})
\end{align*}
\begin{equation}
= 3.05 - 1.95 + 7.09 = 8.19\:\mathrm{eV}
\end{equation}
resulting in a very large enhancement factor
\begin{equation}
\mathrm{log\,}K \sim 90 \, \, \, \mathrm{at} \, T=\mathrm{460\:K}
\end{equation}
The $^{137}$La (6$\times10^4$y) atoms that have been produced in Eq. (25) will undergo a successive reaction
\begin{equation}
^{137}\mathrm{La} + \alpha \rightarrow \, ^{141}\mathrm{Pr}^{3+} + \mathrm{e_{int}^-} \, , \, \mathrm{Q = 1.17\,MeV}
\end{equation}
with the enhancement 
\begin{equation}
\Delta \phi_r^* = \phi^*(\mathrm{Pr})- \phi^*(\mathrm{La}) - \Delta \mathrm{G}_f(\mathrm{Pr^{3+}_{liq}})
\end{equation}
\begin{align*}
\sim 2.83 - 2.75 + 7.0 = 7.08\:\mathrm{eV}
\end{align*}
\begin{equation}
\mathrm{log}\,K \sim 77 \, \, \, \mathrm{at} \, T=\mathrm{460\:K}
\end{equation}
In the cascade chemonuclear reactions Sr$\rightarrow$Zr$\rightarrow$Mo and Cs$\rightarrow$La$\rightarrow$Pr, the accumulation time for Mo/Sr$\rightarrow$1 and Pr/Cs$\rightarrow$1 is governed by the rate enhancement -- i.e., log $K$ $\sim$ 24 in Eq. (24) and log $K$ $\sim$ 77 in Eq. (30). As such, the accumulation time $T$(Mo/Sr$\rightarrow$1) is much longer than that of $T$(Pr/Cs$\rightarrow$1). The experiments of Iwamura et al. [27] reported $T$(Mo/Sr$\rightarrow$1)=300$\sim$400 hrs and $T$(Pr/Cs$\rightarrow$1)$<$100 hrs, giving strong support to our prediction.

\hyphenchar\font=-1
\section*{2.3 Chemonuclear Transmutation}
In the presence of electron donor atoms or molecules, $\gamma$-rays are converted into Auger electrons immediately. As such, we are able to detect $\gamma$-rays that have been produced in the nuclear radiative capture reaction or positron annihilation in a system of reactants and metallike hydride/electron donor mixtures. We specify the relationship
\begin{equation}
\gamma + \mathrm{M} \rightarrow \mathrm{ne^-} + \mathrm{M^{n+}}\, , \, \mathrm{n} = 1,2,3,4
\end{equation}
where M denotes an electron donor atom/molecule. The conversion of Eq. (31) is specified by the chemical potential change $\Delta \phi_r^*(>0)$ and enhanced with the factor exp$[\Delta \phi_r^* / \mathrm{k_B}T]$ due to thermodynamical liquid activity. LiD molecules are dissolved on the surface of the Ni nanocrystal in a mixture of Ni nanopowder and LiD. The Li atoms are adsorbed, resulting in a significant reduction of the work function of the Ni nanocrystal surface according to Eq. (31). In this case, M=Li and M=Ni. We estimate the respective chemical potential changes and enhancement factors as
\begin{equation}
\Delta \phi_r^* = -\Delta \mathrm{G}_f^*(\mathrm{Li_{aq}^+})=3\:\mathrm{eV}, \, \mathrm{log\,}K =33 \,\,\, \mathrm{at}\, T\mathrm{=460\:K}
\end{equation}
and
\begin{equation}
\Delta \phi_r^* = -\Delta \mathrm{G}_f^*(\mathrm{Ni_{aq}^{++}})=0.48\:\mathrm{eV}, \, \mathrm{log\,}K =5.3 \,\,\, \mathrm{at}\, T\mathrm{=460\:K}
\end{equation}
The conversion coefficient is some hundredths for the $\gamma$-rays of around one MeV energy, Eqs. (32) and (33) show that all $\gamma$-rays are converted into electrons that produce a large amount of soft x-rays and phonons. This kind of soft x-rays have been observed in fusion experiments [26]. In the coherent chemonuclear $\mathrm{D_2}$-$\mathrm{D_2}$ and $\mathrm{D_3}$-$\mathrm{D_3}$ fusion, the $\alpha$-particles that have been produced with kinetic energy of 23.85 MeV undergo the resonance capture by Ni nuclei, yielding the intermediate compound Zn$^m$ nuclei at the giant dipole resonance (GDR) states $\hbar\omega_\mathrm{GDR} \sim 20$ MeV with the width $\Gamma \sim 5$ MeV. The reaction cross section summed in that width is as large as
\begin{equation}
\int \sigma\mathrm{d}E \geq 1.5 (\mathrm{MeV.b})
\end{equation}
In Eq. (35), the compound Zn$^m$ nucleus formation will be enhanced  where the atomic and nuclear reactions are spontaneous
\begin{equation}
^\mathrm{A}\mathrm{Ni} + \alpha \rightarrow \, ^\mathrm{A+4}\mathrm{Zn}^{2+}
\end{equation}
This process is governed by the chemical potential change
\begin{align*}
\Delta \phi_r^*(\mathrm{Zn}^{2+}) = \phi^*(\mathrm{Zn})- \phi^*(\mathrm{Ni}) - \Delta \mathrm{G}_f(\mathrm{Zn^{2+}_{liq}})
\end{align*}
\begin{equation}
= 4.10 - 5.20 + 1.52 = 0.42\:\mathrm{eV}
\end{equation}
and enhanced with the factor
\begin{align*}
\mathrm{log}\,K=4.6 \,\,\, \mathrm{at} \, T=\mathrm{460\:K}
\end{align*}
\begin{equation}
\mathrm{log}\,K=7.1 \,\,\, \mathrm{at} \, T=\mathrm{300\:K}
\end{equation}
The enhanced values of the summed reaction cross sections are as follows:
\begin{align*}
\mathrm{6\times10^4(MeV.b) \,\,\, at \, }T=\mathrm{460\:K}
\end{align*}
\begin{equation}
\mathrm{1.8\times10^7(MeV.b) \,\,\, at \, }T=\mathrm{300\:K}
\end{equation}
The values of Eq. (38) are far above the cross section of the thermal neutron capture by $^{235}$U nuclei. The compound nuclei $^{\mathrm{A}+4}$Zn$^m$ will undergo the proton pair emission due to their large decay width. The $^{\mathrm{A}}$Ni nuclei will undergo the 2n and 4n transfer reactions yielding respective $^{\mathrm{A}+2}$Ni and $^{\mathrm{A}+4}$Ni nuclei. The spectroscopical comparison of the radiative neutron capture (n,$\gamma$) reactions and the (d,p) reactions provided us with the first evidence of “Doorway states” [28]. The ground state spin data $^{57}\mathrm{Fe}_{31}$ ($^1/_2$), $^{59}\mathrm{Fe}_{33}$ ($^3/_2$) and $^{61}\mathrm{Ni}_{33}$ ($^3/_2$) indicate the ground state neutron configuration of the Ni isotopes as follows:

\vspace{0.5cm}

\noindent $^{58}\mathrm{Ni}_{30}, (p^3/_2)_2(p^1/_2)_0, U^2(p^1/_2)$ = 1 ; $^{59}\mathrm{Ni}_{31}, (p^3/_2)_2(p^1/_2)_1, U^2(p^1/_2)$ = 1/2

\noindent $^{60}\mathrm{Ni}_{32}, (p^3/_2)_2(p^1/_2)_2, U^2(p^3/_2)$ = 1/2 ; $^{61}\mathrm{Ni}, (p^3/_2)_3(p^1/_2)_2, U^2(p^3/_2)$ = 1/4

\noindent $^{62}\mathrm{Ni_{34}}, (p^3/_2)_4(p^1/_2)_2, U^2(p^3/_2)$ = 0

\vspace{0.5cm}

\noindent Here the neutron hole fraction in the p-shell is indicated, as shown by the non-occupancy fraction $U^2$ of neutron. The structure independent factor of the $2n$-transfer reaction amplitude $\psi$ is common to the Ni isotopes. The product isotope distribution depends only on the natural abundance $\varepsilon$ of the reactant nuclei. The $2n$-transfer reaction amplitude is extremely large, so the reaction takes place except for U$^2$=0. 

The relative yield of each Ni isotope is listed below:

\vspace{0.5cm}

$^{58}\mathrm{Ni} \rightarrow \, ^{60}\mathrm{Ni} \rightarrow \, ^{62}\mathrm{Ni} :$

$\,\,\,\,\,\,\,\,\,\,\, \mathrm{y(^{62}Ni})=\varepsilon(^{58}\mathrm{Ni})\cdot\psi^2U^2(p^1/_2)\cdot\psi^2U^2(p^3/_2)=\varepsilon(^{58}\mathrm{Ni})\cdot\psi^4/2$

$^{60}\mathrm{Ni} \rightarrow \, ^{62}\mathrm{Ni} :$

$\,\,\,\,\,\,\,\,\,\,\, \mathrm{y(^{62}Ni)}=\varepsilon(^{60}\mathrm{Ni})\cdot\psi^2U^2(p^3/_2)=\varepsilon(^{60}\mathrm{Ni})\cdot\psi^2/2$

$^{62}\mathrm{Ni} \rightarrow \, ^{64}\mathrm{Ni} :$

$\,\,\,\,\,\,\,\,\,\,\, \mathrm{y(^{64}Ni})=0\,, \,\,\, \because \, U^2(p^3/_2,p^1/_2)=0 \,$

\noindent The $^{61}\mathrm{Ni(\alpha,2p})$ $^{63}\mathrm{Ni}$ reaction has been neglected, because $U^2$($\,p^3/_2)=\,^1/_4$ means that $^{61}\mathrm{Ni}$ favours the single neutron transfer reaction feeding $^{62}\mathrm{Ni}$. In sum, all Ni isotopes are transmuted into $^{62}\mathrm{Ni}$ except for $^{64}\mathrm{Ni}$.

\hyphenchar\font=-1
\small
\begin{center}
\small
\noindent Table 2.2 \, Measured abundance [\%] for Ni$^+$ ions.
\small
\vspace{0.5 cm}
\begin{tabular}{| c | c | c | c |} \hline
Ion & Fuel & Ash & Theoretical abundance of ash \\ \hline
$^{58}$Ni & 67 & 0.8 & 0\\ \hline
$^{60}$Ni & 26.3 & 0.5 & 0\\ \hline
$^{61}$Ni & 1.9 & 0 & 0\\ \hline
$^{62}$Ni & 3.9 & 98.7 & 100\\ \hline
$^{64}$Ni & 1 & 0 & 0 \\ \hline
\end{tabular}
\end{center}

\noindent 
\normalsize
Assuming that all fuel atoms undergo the reactions, we derived the values of Table 2.2 based on the work of Bexell and Hall [29]. In Table 2.2, the mass analyzing data of 116 hrs chemonuclear fusion running a Ni powder show the above predicted mass shift clearly [29]. Such ($\alpha,2$p) reactions are unlikely for the Pd isotopes.

The chemonuclear fission of non-fertile material (e.g., Pd and Cs isotopes) provides compelling evidence of the cascade chemonuclear reaction in the system of metallike hydride--electron donor mixtures. Contrary to the Ni+$\alpha$ case, the reactions of compound nucleus formation
\begin{equation}
\mathrm{^APd + \alpha \rightarrow \, ^{A+4}Cd^{2+}}
\end{equation}
are not enhanced, because the chemical potential change in the reaction is not spontaneous ($\Delta \phi_r^* < 0$) in that
\begin{align*}
\Delta\phi^*_r(\mathrm{Cd}^{2+})=\phi^*(\mathrm{Cd})-\phi^*(\mathrm{Pd})-\Delta \mathrm{G}_f(\mathrm{Cd^{2+}_{liq}})
\end{align*}
\begin{equation}
=4.05-5.45+0.81=-0.59\:\mathrm{eV}
\end{equation}

In the $\alpha$-capturing chemonuclear reaction, the Pd atoms remain stable, making a strong contrast with the other hydrogen adsorbent Ni and Ti metals. Nonetheless, the chemonuclear fission of Pd nuclei may take place through the GDR states excitation. In the case of the fission of $^{106}$Pd nucleus via the GDR states excitation
\begin{align*}
_{46}^{106}\mathrm{Pd}(\alpha,\alpha')^{106}\mathrm{Pd^{GDR}}\rightarrow\,2\cdot ^{53}\mathrm{V} \,,\, \mathrm{Q=12.1\,MeV}
\end{align*}
\begin{equation}
\hbar\omega_{\mathrm{GDR}}=16.5\,\mathrm{MeV}
\end{equation}

\noindent 
The fissibility parameter of the $^{106}$Pd nucleus is
\begin{equation}
\mathrm{\chi = Z^2/50.12A=0.4}
\end{equation}
The expected spontaneous fission time $T_\mathrm{{SF}}$ is
\begin{equation}
\mathrm{log}\,T_\mathrm{SF}\sim38
\end{equation}
based on the semi-empirical formula
\begin{equation}
\mathrm{log}\,T_\mathrm{SF}\sim -85.0 \, \chi + 72.1
\end{equation}
derived from the fission data on $^{248,250}$Cm, $^{252,254}$Cf, $^{254,256}$Fm and $^{252,256}$No nuclei. The value of Eq. (43) triples those of the $^{238}$U and $^{232}$Th nuclei. The latter values are listed below:
\begin{equation}
\mathrm{^{238}U: \chi=0.71, \,log} \,T_\mathrm{SF}=12
\end{equation}
\begin{align*}
\mathrm{^{232}Th: \chi=0.70, \,log }\,T_\mathrm{SF}=13
\end{align*}
Their induced fission through the inelastically excited GDR states has been observed with some hundreds mb cross sections [30]. 

The induced fission of the Pd nucleus is least likely. Nonetheless, the separation speed of fragment $^{53}$V atoms is 0.016 $c$ (where $c$ is the speed of light), which is far below the gyration speed $Zv_\mathrm{B}=Z\alpha c=0.17\:c$ of their 1s orbital electrons. As such, the electrons are able to adjust their configuration smoothly to match the fission process. The nuclear fission is strictly governed by the thermodynamic force, resulting in the enhanced chemonuclear fission. The rate is governed by the chemical potential change of the chemonuclear fission through the inelastic Pd$^\mathrm{GDR} (\hbar\omega_{\mathrm{GDR}}=16.5$\,MeV) excitation (with the two fragments listed in Table 2.3) as follows:
\begin{equation}
\Delta \phi^*_r = \phi^* \mathrm{(Frag.1)}+\phi^*\mathrm{(Frag.2)}-\phi^*\mathrm{(Pd)}
\end{equation}

\hyphenchar\font=-1
\small
\begin{center}
\small
\noindent Table 2.3 \, Fragments of the Pd fission.
\small
\vspace{0.5 cm}
\begin{tabular}{l c l c l c l c l c l c l} \hline
Fragment 1: & Co & Fe & Mn & Cr & V \\ \hline
Fragment 2: & K & Ca & Sc & Ti & V \\ \hline
$\Delta \phi^*_r\mathrm{(eV)}:$ & 1.90 & 2.03 & 2.25 & 2.85 & 3.05 \\ \hline
log $K$ at $T$ = 460 K & 21 & 22 & 25 & 31 & 33 \\ \hline
\end{tabular}
\end{center}

\normalsize
The enhancement $K$ values suggest that the greater part of the Pd nuclei undergo the $\alpha$-induced chemonuclear fission. Nonetheless, these $K$ values have no bearing on the atomic number distribution of the intrinsic fission fragments due to their $\beta$-decay modes. In the $^{106}$Pd fission, the primary and final product pairs are (at $T$=460 K) listed below:
\begin{equation}
\mathrm{2\cdot\,^{53}V\rightarrow\,2\cdot\,^{53}Cr,\,\,\,\,\,\,\Delta\phi}^*_r\mathrm{=3.05\:eV,\,log\,}K=33
\end{equation}
\begin{equation}
\mathrm{^{55}Cr+\,^{51}Ti\rightarrow\,^{55}Mn+\,^{51}V,\,\,\,\,\Delta\phi}^*_r\mathrm{=2.85\:eV,\,log\,}K=31
\end{equation}
\begin{equation}
\mathrm{^{56}Mn+\,^{50}Sc\rightarrow\,^{56}Fe+\,^{50}Ti,\,\,\,\,\Delta\phi}^*_r\mathrm{=2.25\:eV,\,log\,}K=25
\end{equation}
\begin{equation}
\mathrm{^{60}Fe+\,^{46}Ca\rightarrow\,^{60}Ni+\,^{46}Ca,\,\,\,\,\Delta\phi}^*_r\mathrm{=2.03\:eV,\,log\,}K=22
\end{equation}

\noindent 
As such, we expect the final product distribution to have a central peak of Cr and sub-peaks of Mn and V with some \% yield compared to the Cr peak. Some weak but still visible peaks of Ni, Fe, Ti and Ca may be detected. The $^{104,105,108,110}$Pd fission may result in the central peak of Cr isotopes. These predictions are well supported by the observational data of Mizuno and other groups [31]. Adsorbent isotopes Ni and Ti are an exception, because their nuclear fission is endothermic and their chemonuclear fission is unlikely.

In a system of metallike hydride--electron donor mixtures, $^{137}$Cs undergoes the $\alpha$-capturing chemonuclear reaction (log $K\sim90$ at $T$=460 K), thereby producing $^{141}$La which decays into $^{141}$Pr through the cascade $\beta^-$decays. This radioisotope may also be transmuted into useful elements through the $\alpha$-induced chemonuclear fission via the GDR states excitation
\begin{equation}
\mathrm{^{137}Cs(\alpha,\alpha')\,^{137}Cs^{GDR}\rightarrow\,^{87}Br+\,^{50}Ca,Q=26.7\,MeV}
\end{equation}
provided that \,\,\,\,\,\,\,\,\,\,\,\,\,\,\,\,\,\,\,\,\,\,\,\,\,\,\,\,\,\,\,\,\,$^{87}$Br(55s)$\xrightarrow[\mathrm{6.1\,MeV\beta^-}]{}$ $^{87}$Kr(76m) $\xrightarrow[\mathrm{3.9\,MeV\beta^-}]{}$ $^{87}$Rb
\vspace{0.5cm}

\noindent and \,\,\,\,\,\,\,\,\,\,\,\,\,\,\,\,\,\,\,\,\,\,\,\,\,\,\,\,\,\,\,\,\,\,\,\,\,\,\,\,\,\,\,\,\,\,$^{50}$Ca(9s)$\xrightarrow[\mathrm{3.9\,MeV\beta^-}]{}$ $^{50}$Sc(1.7m) $\xrightarrow[\mathrm{6.5\,MeV\beta^-}]{}$ $^{50}$Ti

\vspace{0.5cm}
\noindent or 

\begin{equation}
\mathrm{^{137}Cs(\alpha,\alpha')\,^{137}Cs^{GDR}\rightarrow\,^{89}Br+\,^{48}Ca,Q=27\,MeV}
\end{equation}
\vspace{0.5cm}

\noindent provided that \,\,\,\,\,\,$^{89}$Br(46s)$\xrightarrow[\mathrm{6\,MeV\beta^-}]{}$ $^{89}$Kr(3.2m) $\xrightarrow[\mathrm{4.6\,MeV\beta^-}]{}$ $^{89}$Rb(15.9m)
\begin{center}
$\xrightarrow[\mathrm{3.9\,MeV\beta^-}]{}$ $^{89}$Sr(52d) $\xrightarrow[\mathrm{1.5\,MeV\beta^-}]{}$ $^{89}$Y
\end{center}

\noindent In the presence of thermodynamical liquid activity, the reactions of Eqs. (52) and (53) are enormously enhanced. The chemical potential change is specified as
\begin{align*}
\Delta\phi^*_r = \phi^*\mathrm{(Br)+\phi^*(Ca)-\phi^*(Cs)}
\end{align*}
\begin{equation}
=7.40 + 2.50 - 1.98 = 7.92\:\mathrm{eV}
\end{equation}
\begin{equation}
\mathrm{log}\,K=87\,\,\,\mathrm{at}\,\,\,T=\mathrm{460\:K} \, 
\end{equation}

The chemonuclear reaction may take place in a system of metallike hydride--electron donor mixtures with nanopowder B or filled with borane gas (B$_2$D$_6$/B$_2$H$_6$). This process is similar to the Big Bang nucleosynthesis (e.g., the $^{11}$B$(\alpha,$n)$\,^{14}$N and $^{11}$B$(\alpha,$p)\,$^{14}$C reactions) that takes place in an inhomogeneous universe. The $\alpha$-particles induce the chemonuclear fusion while reducing their speed at $E$=0.4\,MeV, $v$=$2v_\mathrm{B}$=2$\alpha c$ (where $\alpha$ is the fine structure constant, and $c$ is the speed of light) with the respective cross sections of
\begin{equation}
\mathrm{\sigma(\alpha,n)=6\times10^{-12}}S\mathrm{(\alpha,n)\sim10^{-8}(b),Q=0.16\,MeV}
\end{equation}
and
\begin{equation}
\mathrm{\sigma(\alpha,p)=6\times10^{-12}}S\mathrm{(\alpha,p)\sim10^{-8}(b),Q=0.78\,MeV}
\end{equation}
The $S$-factors are estimated from the data on the reaction $^{9}$Be($\alpha,$n$)^{12}$C [23]. Both reactions proceed through the quasi-N atom formation in the system revealing thermodynamical liquid activity. Their rate enhancement factor is estimated as follows:
\begin{equation}
\Delta\phi^*_r = \mathrm{\phi^*(N) - \phi^*(B)=7.00-4.80=2.2\:eV}
\end{equation}
\begin{equation}
\mathrm{log \,}K=40 \,\,\,\mathrm{ at \,\,\, 300\:K\,\,and\,\,log\,}K=24 \,\,\,\mathrm{at}\,\,\,T=\mathrm{460\:K}
\end{equation}

\noindent 
In this case, protons that have been produced in the $^{11}$B($\alpha,$p)\,$^{14}$C reaction may induce the $^{11}$B(p,$\alpha)^8\mathrm{Be}\rightarrow2\cdot\,^{4}\mathrm{He}$ reactions (in which $Q$=8.68\,MeV) enhanced with factors log $K$ = 11.4 at $T$ = 300 K and log $K$ = 7.5 at $T$=460 K respectively towards the possible cyclic $(\alpha,$p)-(p,$\alpha)$ chemonuclear reactions. See $\phi^*$(Be) in Table 2.1 and $\phi^*$(He) in Chapter 1.

\hyphenchar\font=-1
\section*{2.4 Radioactive Waste Disintegration}
\noindent 
In the presence of thermodynamical liquid activity, the disintegration of radioactive isotopes is enhanced, regardless of the kind of emitting particle or the nature of the microscopic interparticle interactions, as long as $-\Delta$G$_r=\Delta\phi _r \leq 0$ holds in the disintegration [22]. In the $\beta$-disintegration, the daughter atoms are most likely in the ionic state due to the nuclear charge shake-off Auger effect, resulting in the lower chemical potential of the daughter atoms and the great enhancement of the disintegration. 

In the system of metallike hydride--electron donor mixtures, $^{137}$Cs disintegrates rapidly without going through a special transmutation procedure as follows:

\begin{equation}
^{137}\mathrm{Cs} \rightarrow \, ^{137}\mathrm{Ba}^{2+} + \beta ^- + \bar{v}_\mathrm{e} + e^-_A
\end{equation}

\noindent In Eq. (60), the disintegration is governed by the chemical potential change

\begin{align*}
\Delta \phi^*_r(\mathrm{Ba}^{2+})=\phi^* (\mathrm{Ba})-\phi^*(\mathrm{Cs})-\Delta G_f(\mathrm{Ba}^{2+}_\mathrm{liq})
\end{align*}
\begin{equation}
=2.32-1.95+5.81=6.18\:\mathrm{eV}
\end{equation}

\noindent  This results in disintegration rate enhancement with the factor
\begin{equation}
\mathrm{log} \, K = 67 \,\, \mathrm{at}\,\, T=\mathrm{460\:K}
\end{equation}
suggesting the instant disintegration of $^{137}$Cs. However, the speed of the disintegration is determined by the solubility of $^{137}$Cs in the system.

In the same system, the radioisotope $^{90}$Sr disintegrates rapidly without going through a transmutation procedure as follows:

\begin{equation}
^{90}\mathrm{Sr} \rightarrow \, ^{90}\mathrm{Y}^{3+} + \beta ^- + \bar{v}_\mathrm{e} + 2e^-_A \, 
\end{equation}

\noindent Here the number of Auger electrons has been adjusted to guarantee the maximum chemical potential drop in the disintegration

\begin{align*}
\Delta \phi^*_r(\mathrm{Y}^{3+})=\phi^* (\mathrm{Y})-\phi^*(\mathrm{Sr})-\Delta G_f(\mathrm{Y}^{3+}_\mathrm{liq})
\end{align*}
\begin{equation}
=3.20-2.40+7.19=7.99\:\mathrm{eV} \, 
\end{equation}
resulting in a very large enhancement
\begin{equation}
\mathrm{log} \, K = 87 \,\, \mathrm{at}\,\, T=\mathrm{460\:K}
\end{equation}

\noindent The resultant $^{90}$Y$^{3+}$(64h) ions are instantly neutralized in the system filled with itinerary s-electrons, hence the super enhanced chemonuclear disintegration

\begin{equation}
^{90}\mathrm{Y} \rightarrow \, ^{90}\mathrm{Zr}^{2+} + \beta ^- + \bar{v}_\mathrm{e} + e^-_A \, 
\end{equation}
with the chemical potential change
\begin{align*}
\Delta \phi^*_r(\mathrm{Zr}^{2+})=\Delta \phi^* (\mathrm{Zr})-\Delta \phi^*(\mathrm{Y})+\Delta G_f(\mathrm{Zr}^{2+}_\mathrm{liq})
\end{align*}
\begin{equation}
=3.40-3.20+6.50 = 6.7\:\mathrm{eV}\,
\end{equation}
and the enhacement
\begin{equation}
\mathrm{log} \, K = 73 \,\, \mathrm{at}\,\, T=\mathrm{460\:K}
\end{equation}

\noindent Here $\Delta G_f(\mathrm{Zr^{2+}_{liq}})$ has been derived from the chemical data $\Delta G_f(\mathrm{ZrCl_2})$ and $2\Delta G_f(\mathrm{Cl^-_{aq}})$ [18].
The enormous enhancement in Eqs. (65) and (68) suggests the instant cascade disintegration of $^{90}$Sr. The speed of the disintegration is determined by the solubility of $^{90}$Sr in the system. In the system of metallike hydride--electron donor mixtures, adsorbed hydrogen atoms are transformed into metallic states and alloyed [20]. Dense itinerary s-electrons supplied by the electron donor cause a drop in the melting point of metallic hydrogen. This system, featuring thermodynamical liquid activity, provides us with the acting space for the super enhanced D-D chemonuclear fusion. The super enhanced He-induced chemonuclear reactions following the $\mathrm{D_n}$-$\mathrm{D_n}$ fusion may produce 23.8 MeV He ions. This system of metallike hydride--electron donor mixtures raises the possibility of radioactive waste vanishment and the mass synthesis of noble metals. This system will likely be filled with high-speed He atoms, He$^+$ ions and the $\alpha$-particles that have been produced in the coherent $\mathrm{D_2}$-$\mathrm{D_2}$ and $\mathrm{D_3}$-$\mathrm{D_3}$ chemonuclear fusion. We looked into the cascade $\alpha$-capture, as well as the $\alpha$-induced neutron transfer and fission reactions. Given that $^{88}$Sr was transmuted into $^{96}$Mo via $^{92}$Zr formation in the system [27], the radioactive $^{90}$Sr isotope could be transmuted into $^{98}$Mo in the same way. 

\hyphenchar\font=-1
\section*{2.5 Noble Metal Synthesis}
Trans-gold nuclei -- Bi, Pb, Tl and Hg -- are in the region of doubly magic numbers $Z=82$ and $N=126$. They are stable nuclei that will undergo exothermic $\alpha$-cluster emissions with stimulation. Stimulated by $\alpha$-particle irradiation, trans-gold nuclei will emit coherently line-up $\alpha$-clusters. In the 15 MeV excited 0$^+$ state within their natural widths, the wave function of the line-up $\alpha$-clusters (e.g., line-up 4$\alpha$-cluster wave function) overlaps the wave function of $^{16}$O. In the chemonuclear D-D fusion, the 4$\alpha$-cluster reveals the thermodynamical activity of oxygen atoms. In the chemonuclear fusion producing the 23.8 MeV $\alpha$-particles (e.g., a system of hydrogen-Ni nanopowder-Li mixtures), those dispersively charged trans-gold atoms may undergo enhanced chemonuclear $\alpha$-cluster emissions, hence the mass synthesis of noble metals.

The most dense and itinerant electron rich plasma can be realized in the system of transition metals (e.g., Ni and Pd nanopowders) adsorbing hydrogen mixed with dense electron donors (e.g., Li metals or alkaline earth oxides). In the transition metals, adsorbed hydrogen atoms are metallized [20] and their melting point is reduced in the presence of dense itinerant s-electrons. As the super enhanced H-H chemonuclear fusion takes place, intense $\alpha$-particles of 23.8 MeV energy are produced, inducing powerful exothermic chemonuclear reactions (e.g., the chemonuclear Th/U fission or element synthesis). 

This mass synthesis system consists of nanopowder metal and electron donor mixtures enveloped with hydrogen (H$_2$/D$_2$) gas. We use nanopowder hydrogen adsorbent metal mixed with the equimolar weight of Li metal or Li hydride (including LiAlD$_4$) and some tens weight percent of reactant trans-gold element powders. The operation of this system is simple and stable up to 1500\textdegree C where the stability of nanopowder adsorbent metals (e.g., Ni) is no longer guaranteed. Nanopowder Ni adsorbent is the best choice, because other adsorbent metals undergo transmutation and lose their adsorbent characteristics due to the enhanced chemonuclear reaction induced by $\alpha$-particles.

We have run the chemonuclear noble metal synthesis test system of Ni nanopowder and Li mixtures of around one molar weight in some tens bar hydrogen atmosphere for a few weeks. Most of the charged trans-gold elements undergo transmutation, yielding platinum group metals and gold with around 10 MWh power generation. This system can be expanded for noble metal mass synthesis use. From our observations, Ti is transmuted into Cr via the enhanced chemonuclear $\alpha$-particle capturing reaction. As $\Delta \phi^*_\mathrm{r}$(Cd)$=\Delta \phi^*_\mathrm{r}$(Pd+$\alpha$)$<0$, Pd is stable against the $\alpha$-particle capturing chemonuclear reaction, but Pd undergoes the extremely enhanced $\alpha$-particle-induced fission. As $\phi^*_\mathrm{r}$(Zn)$=\phi^*_\mathrm{r}$(Ni+$\alpha$)$<0$, Ni is stable against the $\alpha$-particle capturing chemonuclear reaction. Ni is also stiff against the $\alpha$-particle-induced fission, because the fission of Ni is an endothermic reaction. Ni is subject to the ($\alpha$,2p) -- i.e., quasi-double (d,p) stripping -- reaction through the GDR states excitation induced by $\alpha$-particles of 23.8 MeV energy. It should be noted that most of the $^{58}$Ni atoms (68.3\% abundance) remain because heavier Ni isotopes keep their adsorbent characteristics unchanged.

The wave function of the coherently line-up 3-$\alpha$-cluster is identical to the wave function of $^{12}$C in the 7 MeV excited O$^+$. The 3-$\alpha$-cluster reveals the characteristics of $^{12}$C atoms because of the configurational cooperation of orbital electrons in the $\alpha$-cluster emission
\begin{equation}
\mathrm{Ni}+\alpha \rightarrow \, \mathrm{Cr} ^+  + \, ^{12}\mathrm{C} \,\,
\end{equation}
with the chemical potential change
\begin{equation*}
\Delta \phi _\mathrm{r}^* (\mathrm{Cr+C})= \phi^*(\mathrm{Cr})+\phi^*(\mathrm{C})-\phi^*(\mathrm{Ni})
\end{equation*}
\begin{equation}
= 4.65+6.23-5.20 = 5.68\:\mathrm{eV}\,\, 
\end{equation}
and the enhacement
\begin{equation}
\mathrm{log} \, K \sim 62 \,\, \mathrm{at}\,\, T=\mathrm{460\:K}\, 
\end{equation}

In the chemonuclear synthesis of Au from $^{209}$Bi powder (abundance 100\%), the $^{209}$Bi atoms undergo the reaction

\begin{equation}
^{209}\mathrm{Bi}+\alpha \longrightarrow ^{197}\mathrm{Ir}^{4+}+^{16}\mathrm{O}^{2-} \, ,\, Q=15.8 \,\mathrm{MeV}
\end{equation}
Here, the $Q$-value was estimated based on the atomic mass evaluation [32]. The emerging speed of an excited $^{16}$O ion (i.e., line-up 4$\alpha$-cluster, $Q'$=1.4 MeV) is $v \sim$ 0.01 c (where c is the speed of light). The gyration speed of electrons in the 2s-orbitals of the $^{16}$O atom is 4$v_\mathrm{B}$ = 4$\alpha$c = 0.03 c, and in the 1s-orbitals, 8$v_\mathrm{B}$ = 0.06 c (where $v_\mathrm{B}$ is the Bohr speed, and $\alpha$ is the fine structure constant). The gyration speed of s-orbital electrons is much faster than the emerging speed of the $\alpha$-cluster. As they leave the reactant $^{209}$Bi atom, electrons associated with the cluster adjust their configuration continuously and smoothly to match the $\alpha$-cluster emission process. The cluster is most likely emitted in the form of $^{16}$O atom or ion (See the $\phi^*$ values in Table 2.1) as follows:
\begin{align*}
-\Delta \mathrm{G_r}(\mathrm{Ir}^{4+}+\mathrm{O}^{2-}) > \Delta \phi^*_\mathrm{r}(\mathrm{Ir}+\mathrm{O})
\end{align*}
\begin{equation}
= \phi^*(\mathrm{Ir})+\phi^*(\mathrm{O})-\phi^*(\mathrm{Bi})=5.55+8.50-4.15=9.9\:\mathrm{eV}
\end{equation}
Here the chemical potential of high-speed primary $\alpha$-particles is neglected.

The reaction enhancement is
\begin{equation}
\mathrm{log}\,K\sim109 \,\,\, \mathrm{at}\,\,T=\mathrm{460\:K}
\end{equation}

\noindent The value of Eq. (73) is far larger than the enhancement of the chemonuclear fission $^{209}$Bi $\longrightarrow$ Zr + Tc (e.g., log $K$ = 52 at $T$=460 K) [9]. Most of the $^{209}$Bi atoms dispersed in the system undergo the stimulated $\alpha$-cluster emission and will be transmuted into $^{197}$Ir atoms which decay into $^{197}$Au as follows:
\begin{equation}
^{197}\mathrm{Ir}(\mathrm{7m}) \xlongrightarrow[\mathrm{E_\beta}-=2\,\mathrm{MeV}] \,\, ^{197}\mathrm{Pt}(\mathrm{18h}) \xlongrightarrow[\mathrm{E_\beta}-=0.75\,\mathrm{MeV}] \,\, ^{197}\mathrm{Au} \, 
\end{equation}

The $\alpha$-particles produced by the chemonuclear D-D fusion are extremely energetic (23.8 MeV at maximum), thereby inducing Au synthesis through the adiabatically excited GDR (i.e., the giant dipole 13.1 MeV resonance) states of the $^{209}$Bi nuclei as follows:
\begin{equation}
^{209}\mathrm{Bi}(\alpha,\alpha')^{209}\mathrm{Bi}^{\mathrm{GDR}}\longrightarrow \, ^{197}\mathrm{Ir} + \, ^{12}\mathrm{C}, \,\, Q=8.6\,\mathrm{MeV}
\end{equation}
given that $^{197}$Ir$\rightarrow$$^{197}$Au in Eq. (74). The enhancement is
\begin{align*}
\Delta \phi^*_\mathrm{r}(\mathrm{Ir}+\mathrm{C})=\phi^*(\mathrm{Ir})+\phi^*(\mathrm{C})-\phi^*(\mathrm{Bi})
\end{align*}
\begin{equation}
= 5.55 + 6.23 - 4.15 = 7.63\:\mathrm{eV}
\end{equation}
and
\begin{equation}
\mathrm{log\,}K \sim 84 \,\, \mathrm{at}\,\, T=\mathrm{460\:K}\,
\end{equation}

The synthesis of Os through the $^{209}$Bi GDR states excitation 
\begin{equation}
^{209}\mathrm{Bi}(\alpha,\alpha')^{209}\mathrm{Bi}^{\mathrm{GDR}}\longrightarrow \, ^{192}\mathrm{Re} + \, ^{17}\mathrm{O}, \,\, Q=15.1\,\mathrm{MeV}\,
\end{equation}
takes place in parallel with the reaction of Eq. (75). The chemical potential change and the rate enhancement are as follows:
\begin{align*}
\Delta \phi^*_\mathrm{r}(\mathrm{Re}+\mathrm{O})=\phi^*(\mathrm{Re})+\phi^*(\mathrm{O})-\phi^*(\mathrm{Bi})
\end{align*}
\begin{equation}
= 5.40 + 8.50 - 4.15 = 9.75\:\mathrm{eV}
\end{equation}
\begin{equation}
\mathrm{log\,}K \sim 107 \,\, \mathrm{at}\,\, T=\mathrm{460\:K}\,
\end{equation}
The resultant $^{192}$Re atoms will decay into $^{192}$Os as follows:
\begin{equation}
^{192}\mathrm{Re}(16\mathrm{s}) \xlongrightarrow[E_\beta-=2.5\,\mathrm{MeV}]{} \,^{192}\mathrm{Os} \,
\end{equation}

In the chemonuclear synthesis of Au from $^{205}$Tl (70.5\% abundance), the $^{205}$Tl atoms undergo the reaction
\begin{equation}
^{205}\mathrm{Tl} + \alpha \longrightarrow \, ^{197}\mathrm{Ir} + \, ^{12}\mathrm{C}^{2+}, \, \, Q=5.5\,\mathrm{MeV}
\end{equation}
The emerging speed of the 7 MeV excited state $^{12}$C atoms (i.e., line-up 3$\alpha$-clusters) is 0.015 $c$, where $c$ is the speed of light.

\noindent The chemical potential change in the reaction of Eq. (82) is
\begin{align*}
-\Delta\mathrm{G_r}(\mathrm{Ir}+\mathrm{C}^{2+})>\Delta\phi^*_\mathrm{r}(\mathrm{Ir}+\mathrm{C})
\end{align*}
\begin{equation}
=\phi^*(\mathrm{Ir})+\phi^*(\mathrm{C})-\phi^*(\mathrm{Tl})=5.55+6.23-3.90=7.88\:\mathrm{eV}
\end{equation}
The enhancement factor is 
\begin{equation}
\mathrm{log\,}K>86\,\mathrm{at} \,T=\mathrm{460\:K}\,
\end{equation}
We expect that most of the $^{205}$Tl atoms dispersed in this system will be transmuted into $^{197}$Ir.

In the chemonuclear synthesis of Pt and Ir from $^{203}$Tl (29.5\% abundance), the $^{203}$Tl atoms undergo the reaction
\begin{equation}
^{203}\mathrm{Tl}+\alpha \longrightarrow \, ^{195}\mathrm{Ir}+\,^{12}\mathrm{C}^{2+}, \,\, Q=7.7\,\mathrm{MeV}\,
\end{equation}
specified by the chemical potential change $-\Delta\mathrm{G_r}>7.88\:\mathrm{eV}$ in Eq. (83) and the enhancement log $K$ $>86$ in Eq. (84). 
The resultant $^{195}$Ir atoms will decay into $^{195}$Pt as follows:
\begin{equation}
^{195}\mathrm{Ir}(4.2\mathrm{h})\xlongrightarrow[E_\beta-=1\,\mathrm{MeV}]{}\,^{195}\mathrm{Pt} \,
\end{equation}

\noindent However, $^{203}$Tl might be transmuted into $^{191}$Ir as follows:
\begin{equation}
^{203}\mathrm{Tl}+\alpha \longrightarrow \,^{191}\mathrm{Re}^{4+}+\,^{16}\mathrm{O}^{2-}, \,\, Q=14.4\,\mathrm{MeV}\,
\end{equation}
subject to the chemical potential change
\begin{align*}
-\Delta\mathrm{G_r}(\mathrm{Re}^{4+}+\mathrm{O}^{2-})>\Delta\phi^*_\mathrm{r}(\mathrm{Re}+\mathrm{O})
\end{align*}
\begin{equation}
=\phi^*(\mathrm{Re})+\phi^*(\mathrm{O})-\phi^*(\mathrm{Tl})=5.40+8.50-3.90=10\:\mathrm{eV}
\end{equation}
and enhanced enormously
\begin{equation}
\mathrm{log\,}K\sim110\,
\end{equation}
resulting in the production of $^{191}$Ir through Eq. (90), that is
\begin{equation}
^{191}\mathrm{Re}(\mathrm{9.8m}) \xlongrightarrow[\mathrm{E_\beta}-=1.8 \, \mathrm{MeV}] \,\, ^{191}\mathrm{Os}(\mathrm{15d}) \xlongrightarrow[\mathrm{E_\beta}-=0.31\,\mathrm{MeV}] \,\, ^{191}\mathrm{Ir} \,
\end{equation}

The $^{191}$Ir synthesis may also take place through the GDR ($\hbar\omega_\mathrm{GDR}=13.3$ MeV) states excitation of the $^{203}$Tl nuclei as follows:
\begin{equation}
^{203}\mathrm{Tl}(\alpha,\alpha')\,^{203}\mathrm{Tl}^{\mathrm{GDR}}\longrightarrow\,^{191}\mathrm{Re}+\,^{12}\mathrm{C},\,\,Q=1.24\,\mathrm{MeV}\,
\end{equation}
\begin{equation}
\Delta\phi^*_\mathrm{r}(\mathrm{Re}+\mathrm{C})=7.73\:\mathrm{eV},\,\, \mathrm{log\,}K=93\,
\end{equation}

Lead isotopes are transmuted into Au, Pt and Ir through the stimulated emissions of the 3$\alpha$ or 4$\alpha$-clusters (or their neighboring clusters).

In the chemonuclear synthesis of Pt and Au from $^{208}$Pb (52.3\% abundance), most of the $^{208}$Pb atoms undergo the reaction
\begin{equation}
^{208}\mathrm{Pb}+\alpha \longrightarrow \,^{194}\mathrm{Os}^{4+}+\,^{18}\mathrm{O}^{2-}, \,\, Q=12.3\,\mathrm{MeV}\,
\end{equation}
with the enhancement
\begin{align*}
-\Delta\mathrm{G_r}(\mathrm{Os}^{4+}+\mathrm{O}^{2-})>\Delta\phi^*_\mathrm{r}(\mathrm{Os}+\mathrm{O})
\end{align*}
\begin{equation}
=\phi^*(\mathrm{Os})+\phi^*(\mathrm{O})-\phi^*(\mathrm{Pb})=5.40+8.50-4.10=9.8\:\mathrm{eV}
\end{equation}
\begin{equation}
\mathrm{log\,}K\sim107\,\mathrm{at}\,\,T=\mathrm{460\:K}\,
\end{equation}
The production of the $^{194}$Os nuclei may also take place through the GDR states excitation of the $^{208}$Pb nuclei as follows:
\begin{equation}
^{208}\mathrm{Pb}(\alpha,\alpha')\,^{208}\mathrm{Pb}^{\mathrm{GDR}}\longrightarrow\,^{194}\mathrm{Os}+\,^{14}\mathrm{C},\,\,Q=6.07\,\mathrm{MeV}\,
\end{equation}
\begin{equation}
\Delta\phi^*_\mathrm{r}=5.40+6.23-4.10=7.53\:\mathrm{eV}
\end{equation}
\begin{equation}
\mathrm{log\,}K=83\,\,\mathrm{at}\,\,T=\mathrm{460\:K}\,
\end{equation}
The resultant $^{194}$Os nuclei will decay into $^{194}$Pt via $^{194}$Ir as follows: 
\begin{equation}
^{194}\mathrm{Os}(\mathrm{6.0y}) \xlongrightarrow[\mathrm{E_\beta}-=0.097 \, \mathrm{MeV}] \,\, ^{194}\mathrm{Ir}(\mathrm{17.4h}) \xlongrightarrow[\mathrm{E_\beta}-=2.34 \, \mathrm{MeV}] \,\, ^{194}\mathrm{Pt} \,
\end{equation}

\noindent However, a smaller but non-vanishing part of the $^{208}$Pb atoms might undergo the reaction
\begin{equation}
^{208}\mathrm{Pb}+\alpha \longrightarrow \,^{197}\mathrm{Ir}+\,^{15}\mathrm{N}^{2+}, \,\, Q=7.4\,\mathrm{MeV}\,
\end{equation}
resulting in Au synthesis with the enhancement
\begin{align*}
-\Delta\mathrm{G_r}(\mathrm{Ir}+\mathrm{N}^{2+})>\Delta\phi^*_\mathrm{r}(\mathrm{Ir}+\mathrm{N})
\end{align*}
\begin{equation}
=\phi^*(\mathrm{Ir})+\phi^*(\mathrm{N})-\phi^*(\mathrm{Pb})=5.55+7.00-4.10=8.45\:\mathrm{eV}
\end{equation}
\begin{equation}
\mathrm{log\,}K>93\,\mathrm{at}\,\,T=\mathrm{460\:K}\,
\end{equation}
The resultant $^{197}$Ir nuclei will decay into $^{197}$Au based on Eq. (74).

In the chmeonuclear synthesis of Ir, Pt and Au from $^{207}$Pb (22.6\% abundance), a large part of the $^{207}$Pb atoms are transmuted into $^{193}$Ir and $^{195}$Pt with an approximately equal probability as follows:
\begin{equation}
^{207}\mathrm{Pb}+\alpha\longrightarrow\,^{193}\mathrm{Os}^{4+}+\,^{18}\mathrm{O}^{2-}\,,\,\,Q=13.3\,\mathrm{MeV}\,
\end{equation}
\begin{equation}
^{207}\mathrm{Pb}+\alpha\longrightarrow\,^{195}\mathrm{Os}^{4+}+\,^{16}\mathrm{O}^{2-}\,,\,\,Q=13\,\mathrm{MeV}\,
\end{equation}
Both reactions are enhanced with the factor $K \sim107$ based on Eq. (95).

The production of the $^{193,195}$Os nuclei may also take place through the GDR states excitation of the $^{207}$Pb nuclei as follows:
\begin{equation}
^{207}\mathrm{Pb}(\alpha,\alpha')^{207}\mathrm{Pb}^{\mathrm{GDR}}\longrightarrow\,^{193}\mathrm{Os}+\,^{14}\mathrm{C}\,,\,\, Q=7.12\,\mathrm{MeV}\,
\end{equation}
\begin{equation}
^{207}\mathrm{Pb}(\alpha,\alpha')^{207}\mathrm{Pb}^{\mathrm{GDR}}\longrightarrow\,^{195}\mathrm{Os}+\,^{12}\mathrm{C}\,,\,\, Q=5.78\,\mathrm{MeV}\,
\end{equation}
with the enhancement log $K = 83$ in Eq. (97).

The resultant $^{193}$Os and $^{195}$Os nuclei will decay as follows:
\begin{equation}
^{193}\mathrm{Os}(31\mathrm{h})\xlongrightarrow[\mathrm{E_\beta}-=1.13 \, \mathrm{MeV}]{} \,^{193}\mathrm{Ir}\,
\end{equation}
\begin{equation}
^{195}\mathrm{Os}(6.5\mathrm{m})\xlongrightarrow[\mathrm{E_\beta}-=2\,\mathrm{MeV}]{} \,^{195}\mathrm{Ir}(4.2\mathrm{h})\,\xlongrightarrow[\mathrm{E_\beta}-=1\,\mathrm{MeV}]{} \,^{195}\mathrm{Pt}\,
\end{equation}

One part of the $^{207}$Pb nuclei might be transmuted into $^{197}$Au through the production of $^{197}$Ir based on Eq. (74) as follows:
\begin{equation}
^{207}\mathrm{Pb}+\alpha\longrightarrow\,^{197}\mathrm{Ir}+\,^{14}\mathrm{N}^{2+}\,,\,\,Q=4\,\mathrm{MeV}\,
\end{equation}
with the enhancement log $K>93$ in Eq. (102).

Another part of the $^{207}$Pb nuclei might be transmuted into $^{197}$Au through the production of $^{197}$Pt based on Eq. (74) as follows:
\begin{equation}
^{207}\mathrm{Pb}+\alpha\longrightarrow\,^{197}\mathrm{Pt}+\,^{14}\mathrm{C}^{2+}\,,\,\,Q=6.6\,\mathrm{MeV}\,
\end{equation}
with the enhancement
\begin{align*}
-\Delta\mathrm{G_r}(\mathrm{Pt}+\mathrm{C}^{2+})>\Delta\phi^*_\mathrm{r}(\mathrm{Pt}+\mathrm{C})
\end{align*}
\begin{equation}
=\phi^*(\mathrm{Pt})+\phi^*(\mathrm{C})-\phi^*(\mathrm{Pb})=5.65+6.23-4.10=7.78\:\mathrm{eV}
\end{equation}
\begin{equation}
\mathrm{log\,}K>85\,\mathrm{at}\,\,T=\mathrm{460\:K}\,
\end{equation}

In the chemonuclear synthesis of Au and Pt from $^{206}$Pb (23.6\% abundance), the $^{13}$N and $^{13}$C nuclei are near 3$\alpha$-cluster nuclei with respective configurations of 3$\alpha+p$ and 3$\alpha+n$. Their binding energies are fairly large (like those of $^{14}$C and $^{14}$N), and they tend to be emitted from $^{206}$Pb due to $\alpha$-particle irradiation, resulting in the Au synthesis below 

\begin{equation}
^{206}\mathrm{Pb}+\alpha\longrightarrow\,^{197}\mathrm{Ir}+\,^{13}\mathrm{N}^{2+}\,,\,\,Q=2\,\mathrm{MeV}\,
\end{equation}
with the enhancement log $K>93$ based on Eq. (102)
\begin{equation}
^{206}\mathrm{Pb}+\alpha\longrightarrow\,^{197}\mathrm{Pt}+\,^{13}\mathrm{C}^{2+}\,,\,\,Q=5.2\,\mathrm{MeV}\,
\end{equation}
and with the enhancement log $K>85$ based on Eq. (112). 

The resultant $^{197}$Ir and $^{197}$Pt nuclei will decay into $^{197}$Au based on Eq. (74). The remaining part of the $^{206}$Pb nuclei are transmuted into $^{198}$Pt
\begin{equation}
^{206}\mathrm{Pb}+\alpha\longrightarrow\,^{198}\mathrm{Pt}+\,^{12}\mathrm{C}^{2+}\,,\,\,Q=8.6\,\mathrm{MeV}\,
\end{equation}
with the enhancement log $K > 85$ based on Eq. (112).

In the chemonuclear synthesis of Pt, Ir and Os from Hg, all Hg isotopes undergo the stimulated C ion emitting reaction
\begin{equation}
\mathrm{Hg}+\alpha\longrightarrow\mathrm{Os}+\mathrm{C}^{2+}\,
\end{equation}
specified by the chemical potential change
\begin{align*}
-\Delta\mathrm{G_r}(\mathrm{Os}+\mathrm{C}^{2+})>\Delta\phi^*_\mathrm{r}(\mathrm{Os}+\mathrm{C})
\end{align*}
\begin{equation}
=\phi^*(\mathrm{Os})+\phi^*(\mathrm{C})-\phi^*(\mathrm{Hg})=5.40+6.23-4.20=7.43\:\mathrm{eV}\,
\end{equation}
and enhanced with the factor
\begin{equation}
\mathrm{log}\,K>82\,\,\mathrm{at}\,\,T=\mathrm{460\:K}\,
\end{equation}

In the chemonuclear synthesis of Pt from $^{204}$Hg (6.85\% abundance), the enhanced reaction (log $K>82$) due to $\alpha$-particle irradiation 
\begin{equation}
^{204}\mathrm{Hg}+\alpha\longrightarrow\,^{195}\mathrm{Os}+\,^{13}\mathrm{C}^{2+}\,,\,\,Q=2.8\,\mathrm{MeV}\,
\end{equation}
takes place. The resultant $^{195}$Os nuclei will decay into $^{195}$Pt via $^{195}$Ir, that is
\begin{equation}
^{195}\mathrm{Os}(6.5\mathrm{m})\xlongrightarrow[\mathrm{E_\beta}-=2\,\mathrm{MeV}]{} \,^{195}\mathrm{Ir}(4.2\mathrm{h})\,\xlongrightarrow[\mathrm{E_\beta}-=1\,\mathrm{MeV}]{} \,^{195}\mathrm{Pt}\,
\end{equation}

In the chemonuclear synthesis of Pt and Ir from $^{202}$Hg (29.8\% abundance), the enhanced reaction (log $K>82$)
\begin{equation}
^{202}\mathrm{Hg}+\alpha\longrightarrow\,^{194}\mathrm{Os}+\,^{12}\mathrm{C}^{2+}\,,\,\,Q=5.9\,\mathrm{MeV}\,
\end{equation}
produces $^{194}$Pt, that is
\begin{equation}
^{194}\mathrm{Os}(6.0\mathrm{y})\xlongrightarrow[\mathrm{E_\beta}-=0.097 \, \mathrm{MeV}]{} \,^{194}\mathrm{Ir}(17.4\mathrm{h})\,\xlongrightarrow[\mathrm{E_\beta}-=2.29 \, \mathrm{MeV}]{} \,^{194}\mathrm{Pt}\,
\end{equation}

A part of the $^{202}$Hg nuclei will undergo the reaction (log $K>82$)
\begin{equation}
^{202}\mathrm{Hg}+\alpha\longrightarrow\,^{193}\mathrm{Os}+\,^{13}\mathrm{C}^{2+}\,,\,\,Q=5.4\,\mathrm{MeV}\,
\end{equation}
resulting in $^{193}$Ir synthesis based on Eq. (107).

In the chemonuclear synthesis of Ir and Os from $^{201}$Hg (13.2\% abundance), the enhanced reaction (log $K>82$)
\begin{equation}
^{201}\mathrm{Hg}+\alpha\longrightarrow\,^{193}\mathrm{Os}+\,^{12}\mathrm{C}^{2+}\,,\,\,Q=8.2\,\mathrm{MeV}\,
\end{equation}
results in $^{193}$Ir synthesis, which is similar to Eq. (123).

A part of the $^{201}$Hg nuclei will also undergo the Os synthesis reaction (log $K>82$) as follows:
\begin{equation}
^{201}\mathrm{Hg}+\alpha\longrightarrow\,^{192}\mathrm{Os}+\,^{13}\mathrm{C}^{2+}\,,\,\,Q=7.5\,\mathrm{MeV}\,
\end{equation}

In the chemonuclear synthesis of Os from $^{200}$Hg (23.1\% abundance), $^{198}$Hg(10.0\%) and $^{196}$Hg(0.15\%), the enhanced reaction (log $K>82$)
\begin{equation}
^{200}\mathrm{Hg}+\alpha\longrightarrow\,^{192}\mathrm{Os}+\,^{12}\mathrm{C}^{2+}\,,\,\,Q=8.8\,\mathrm{MeV}\,
\end{equation}
\begin{equation}
^{198}\mathrm{Hg}+\alpha\longrightarrow\,^{190}\mathrm{Os}+\,^{12}\mathrm{C}^{2+}\,,\,\,Q=10.1\,\mathrm{MeV}\,
\end{equation}
\begin{equation}
^{196}\mathrm{Hg}+\alpha\longrightarrow\,^{188}\mathrm{Os}+\,^{12}\mathrm{C}^{2+}\,,\,\,Q=11.7\,\mathrm{MeV}\,
\end{equation}
produces stable Os isotopes.

In the chemonuclear synthesis of Ir from $^{200}$Hg and $^{199}$Hg (16.8\% abundance), the enhanced reaction (log $K>82$)
\begin{equation}
^{200}\mathrm{Hg}+\alpha\longrightarrow\,^{191}\mathrm{Os}+\,^{13}\mathrm{C}^{2+}\,,\,\,Q=5.4\,\mathrm{MeV}\,
\end{equation}
\begin{equation}
^{199}\mathrm{Hg}+\alpha\longrightarrow\,^{191}\mathrm{Os}+\,^{12}\mathrm{C}^{2+}\,,\,\,Q=8.5\,\mathrm{MeV}\,
\end{equation}
produces $^{191}$Ir isotope, that is
\begin{equation}
^{191}\mathrm{Os}(15\mathrm{d}) \xlongrightarrow[\mathrm{E_\beta}-=0.31 \, \mathrm{MeV}]{} \, ^{191}\mathrm{Ir}\,
\end{equation}

To sum up, almost all Bi atoms are transmuted into Au and Os through chemonuclear synthesis. The $^{205}$Tl atoms have mostly been transmuted into Au. The $^{203}$Tl atoms are converted into Pt. Most of the Pb atoms proceed to the Au, Pt and Ir synthesis. All Hg atoms are transmuted into Pt, Ir and Os with an approximately equal probability. As for the Os and Pt synthesis, the chemonuclear $\alpha$-particle capturing of W is the alternative.

\hyphenchar\font=-1
\section*{2.6 Chemonuclear Pion Production}

In a system of Li and Ni hydride nanopowders filled with hydrogen (D$_2$/H$_2$) or borane (B$_2$D$_6$/B$_2$H$_6$) gas, we may well expect the collective production of charged pions and muons that may further cause pionic atom/molecule condensation. At $T\sim500$ K, this system reveals thermodynamical liquid activity induced by dense itinerant s-electrons. We introduce high-energy electrons or photons of some GeV into the system to excite the (3,3) resonance states of component hadrons in the $^7$Li atoms adiabatically without causing any thermodynamical disturbance.

In the presence of thermodynamical liquid activity, the charged pion emissions from the hadrons that we specified below

\begin{equation}
^7\mathrm{Li} \rightarrow \, ^7\mathrm{Be}^{2+} + \pi^- + e^-_A\, 
\end{equation}
\noindent
are governed by the chemical potential change

\begin{align*}
\Delta \phi^*_r=\phi^* (\mathrm{Be})-\phi^*(\mathrm{Li})-\Delta G_f(\mathrm{Be}^{2+}_\mathrm{liq})
\end{align*}
\begin{equation}
=4.20-2.85+3.94=5.29\:\mathrm{eV}
\end{equation}

\noindent Here $e^-_A$ denotes a nuclear charge shake-off Auger electron associated with the charged pion emission.

The pion emissions are enhanced as follows:

\begin{equation}
\mathrm{log}\, K = \mathrm{log}[\mathrm{exp}(-\Delta \phi^*_r/k_\mathrm{B}T] = 58, \,\, \mathrm{at}\, T=\mathrm{460\:K}\,
\end{equation}
\noindent There is little chance of $\pi^0/\pi^+$ emissions and pionization, because a drop in the chemical potential is not expected.

The reactant Li atoms are strongly correlated on macroscopic scale in this system. They react with high-energy electrons/photons coherently, producing low-energy negative pions, hence the largest summed chemical potential drop with the passing through high-energy electrons/photons (e.g. wiggler generating coherent SR photons). The resultant negative pions lead to the formation of pionic atoms. In the presence of thermodynamical liquid activity in this system, pionic atoms are most likely to form covalent pionic molecules. The pionic atoms/molecules will decay into condensed muonic atoms/molecules through the disintegration

\begin{equation}
\pi^- \rightarrow \mu^- + \nu_\mu\,
\end{equation}

\noindent
The size of pionic atoms differs from that of muonic atoms, hence the different chemical potential of pionic and muonic atoms. In this case, the rate of disintegration is governed by this chemical potential difference. If we introduce a heavy element of atomic number Z$>$10, the disintegration of condensed muonic atoms is most likely to take place through the muon capture by their component protons as follows:

\begin{equation}
\mu^- + \mathrm{p} \rightarrow \mathrm{n} + \nu_\mu\,\,
\end{equation}

In this disintegration, most of the liberated energy $m_\mu c^2=106$ MeV has been carried away by the neutrino $\nu_\mu$. The remaining energy of around 30 MeV goes to the residual nucleus, thereby exciting the GDR states. In this case, the disintegration of muonic atoms is associated with the fluctuation in the speed of disintegration.

\small
\section*{References}
\begin{enumerate}
\item Jancovici B (1977) Pair correlation function in a dense plasma and pycnonuclear reactions in stars. J Stat Phys 17:357--370
\item Ichimaru S, Kitamura H (1996) Thermodynamic enhancement of nuclear reactions in dense stellar matter. Publ Astron Soc Jpn 48:613--618
\item Ichimaru S (2000) Radiative proton-capture reactions of high-Z nuclei in the sun and in liquid--metallic hydrogen. Phys Lett A 266:167--172
\item Caughlan GR, Fowler WA (1988) Thermonuclear reaction rates V. Atomic Data and Nuclear Data Tables 40:283--334
\item Fowler WA, Caughlan GR, Zimmerman BA (1967) Thermonuclear reaction rates. Ann Rev Astron Astrophys 5:523--570
\item Cameron AGW (1959) Pycnonuclear reactions and nova explosions. Astrophys J 130:916--940
\item Ikegami H (2001) Buffer energy nuclear fusion. Jpn J Appl Phys 40:6092--6098
\item Ikegami H, Pettersson R (2002) Evidence of enhanced nonthermal nuclear fusion. Bulletin of Institute of Chemistry, Uppsala University
\item Ibid. pp 16--29
\item Ibid. pp 31--41
\item Ikegami H, Pettersson R, Einarsson L (2004) Enormous entropy enhancement revealed in linked nuclear and atomic Li + D fusion in metallic Li liquid. Prog Theo Phys Suppl 154:251--260
\item Ikegami H, Watanabe T, Petterson R, Fransson K (2007) Ultradense nuclear fusion in metallic lithium liquid. Revision of the Swedish Energy Agency Document ER2006:42, Tokyo, pp 1-1--1-22
\item Ibid. pp 2-1--2-12
\item Ibid. pp 3-1--3-30
\item Ibid. pp 4-1--4-21
\item Boom R, de Boer FR, Miedema AR (1976) On the heat of mixing of liquid alloys--II. J Less-Common Metals 46:271--284
\item Pauling L (1960) The nature of the chemical bond, 3rd edn. Cornell University Press, Ithaca
\item Aylward G, Findlay T (1998) SI chemical data, 3rd edn. Wiley, Brisbane
\item Kaufmann EN, Vianden R, Chelikowsky JR, Phillips, JC (1977) Extension of equilibrium formation criteria to metastable microalloys. Phys Rev Lett 39:1671--1674
\item Miedema AR (1973) The electronegativity parameter for transition metals: heat of formation and charge transfer in alloys. J Less-Common Metals 32:117--136
\item Widom B (1963) Some topics in the theory of fluids. J Chem Phys 39:2808--2812
\item Kondepudi D, Prigogine I (1998) Modern thermodynamics. Wiley, Chichester
\item Angulo C et al (1999) A compilation of charged-particle induced thermonuclear reaction rates. Nucl Phys A 656:3--183
\item Fukai Y (2005) The metal--hydrogen system: basic bulk properties, 2nd edn. Springer, Berlin Heidelberg
\item Fukai Y, Sugimoto H (1985) Diffusion of hydrogen in metals. Adv in Phys 34:263--326 
\item Iwamura Y, Itoh T, Yamazaki N, Yonemura H, Fukutani K, Sekiba D (2013) Recent advances in deuterium permeation transmutation experiments. J Condensed Matter Nucl Sci:10:63--71
\item Iwamura Y, Sakano M, Itoh T (2002) Elemental analysis of Pd complexes: effects of D$_2$ gas permeation. Jpn J. Appl Phys 41:4642--4650
\item Ikegami H, Emery GT (1964) Evidence for the excitation of two-particle, one-hole configurations in the capture of thermal neutrons. Phys Rev Lett 13:26--29
\item Levi G, Foschiet E, Höistad B, Pettersson R, Tegnér L, Essén H (2014) Observation of abundant heat production from a reactor device and of isotopic changes in the fuel. Bologna, Uppsala and Stockholm
\item Hicks HG, Levy HB, Nervik WE, Stevenson PC, Niday JB, Armstrong JC (1962) Further radiochemical studies of fission of the ${\mathrm{U}}^{236*}$ compound nucleus. Phys Rev 128:700--707
\item Mizuno T (2000) Experimental confirmation of the nuclear reaction at lower temperature. In: Proceedings for the symposium on advanced research in energy technology 2000, Hokkaido University. pp 95--106
\item Gove NB, Wapstra AH (1972) The 1971 atomic mass evaluation in five parts. Atomic Data and Nuclear Data Tables 11:127--128
\end{enumerate}
\end{document}